\RequirePackage{lineno}
\documentclass[twocolumn,showpacs,superscriptaddress,amsmath,amssymb,nofootinbib]{revtex4-2}
\usepackage{graphicx}
\usepackage{subfigure}
\usepackage{subcaption}
\usepackage{epsfig}
\usepackage{overpic}
\usepackage{dcolumn}
\usepackage{ulem}
\usepackage{bm}
\usepackage{color}
\usepackage{lineno}
\usepackage{xspace}
\usepackage{multirow}
\usepackage{epstopdf}
\usepackage{xcolor}
\usepackage{soul}
\usepackage{verbatim}
\usepackage{enumitem}
\usepackage{todonotes}
\usepackage{cancel}
\usepackage[toc,page]{appendix}
\usepackage[colorlinks,allcolors=black]{hyperref}

\usepackage{diagbox}

  \newcommand{\moe}{\affiliation{Key Laboratory of Atomic and Subatomic Structure and Quantum Control (MOE), Guangdong Basic Research Center of Excellence for Structure and Fundamental Interactions of Matter, Institute of Quantum Matter, South China Normal University, Guangzhou 510006, China
}}

\newcommand{\iqm}{\affiliation{Guangdong-Hong Kong Joint Laboratory of Quantum Matter, Guangdong Provincial Key Laboratory of Nuclear Science, Southern Nuclear Science Computing Center, South China Normal University, Guangzhou 510006, China}}

\newcommand{\scnt}{\affiliation{Southern Center for Nuclear-Science Theory (SCNT), Institute of Modern Physics, Chinese Academy of Sciences, Huizhou 516000, Guangdong Province, China}}

\newcommand{\gscas}{\affiliation{Graduate School of China Academy of Engineering Physics, Beijing 100193, China}}

\begin{document}
\include{def-com}
\title{\boldmath Binding of the three-hadron $DD^*K$ system from the lattice effective field theory}



\author {Zhenyu Zhang}
\iqm
\moe
\author {Xin-Yue Hu}
\iqm
\moe
\author {Guangzhao He}
\iqm
\moe
\author {Jun Liu}
\gscas
\author {Jia-Ai Shi}
\gscas
\author {Bing-Nan Lu}
\email{bnlv@gscaep.ac.cn, corresponding author}
\gscas
\author {Qian Wang}
\email{qianwang@m.scnu.edu.cn, corresponding author}
\iqm
\moe
\scnt
 
\date{\today}

\begin{abstract}

We employ the nuclear lattice effective field theory (NLEFT), an efficient tool for nuclear {\it ab initio} calculations,  to solve the asymmetric multihadron systems. We take the $DD^*K$ three-body system as an illustration to demonstrate the capability of the method. Here the two-body chiral interactions between $D$, $D^*$, and $K$ are regulated with a soft lattice regulator and calibrated with the binding energies of the $T_{cc}^+$, $D^{\ast}_{s0}(2317)$, and $D_{s1}(2460)$ molecular states. 
We then calculate the three-body binding energy using the NLEFT and analyze the systematic uncertainties due to the finite volume effects, the sliding cutoff, and the leading-order three-body forces.
Even when the three-body interaction is repulsive (even as large as the infinite repulsive interaction), the three-body system has a bound state unambiguously with binding energy no larger than the $D_{s1}(2460)D$ threshold. To check the renormalization group invariance of our framework, we extract the first excited state. We find that when the ground state is fixed, the first excited states with various cutoffs coincide with each other when the cubic size goes larger. In addition, the standard angular momentum and parity projection technique is implemented for the quantum numbers of the ground and excited states. We find that both of them are $S$-wave states with quantum number $J^{P}=1^-$. Because the three-body state contains two charm quarks, it is easier to be detected in the Large Hadron Collider.  

\end{abstract}
\maketitle


\section{Introduction}
\label{sec:Introduction}

The color confinement property of 
quantum chromodynamics (QCD), the fundamental theory of strong interaction, allows for the existence of any color singlet object, 
even for the hadrons beyond the conventional quark model. With the increasing energy and statistic in experiment, various experimental collaborations have measured tens of exotic candidates~\cite{Olsen:2014qna}. More specifically, numerous exotic candidates are with charm quark(s), 
for instance the famous $X(3872)$~\cite{Belle:2003nnu}, the recently observed double charm tetraquark $T_{cc}^+$~\cite{LHCb:2021vvq,LHCb:2021auc}, the scalar $D^\ast_{s0}(2317)$~\cite{Belle:2003guh,CLEO:2003ggt}, the axial vector $D_{s1}(2460)$~\cite{Belle:2003guh,CLEO:2003ggt}, the hidden charm pentaquark $P_c$~\cite{LHCb:2015yax, LHCb:2016ztz, LHCb:2019kea}. and the fully charmed tetraquark $X(6900)$~\cite{LHCb:2020bwg, CMS:2023owd, ATLAS:2023bft}.   
To understand the property of these exotic candidates, tremendous theoretical efforts have been put forward, such as compact tetraquarks~\cite{Esposito:2014rxa,Lebed:2016hpi}, hadronic molecules~\cite{Guo:2017jvc}, hybrids~\cite{Meyer:2015eta}, normal heavy quarkonium~\cite{QuarkoniumWorkingGroup:2004kpm}, and so on.  For detailed reviews, one can refer to Refs.~\cite{Esposito:2014rxa,Lebed:2016hpi,Guo:2017jvc,Chen:2016qju,Yamaguchi:2019vea,Brambilla:2019esw,Guo:2019twa,Chen:2022asf,Kalashnikova:2018vkv,Meyer:2015eta}. 

The hadronic molecular picture is proposed based on the fact that most of the exotic candidates locate close to some $S$-wave thresholds. For instance, the $T_{cc}^+$, $D_{s0}^*(2317)$, and $D_{s1}(2460)$ are close to the $DD^*$, $DK$, and $D^*K$ thresholds, respectively. They could be analogous to the deuteron which is a bound state of one proton and one neutron. In this case, the concepts and the methods in nuclear physics can be employed to understand hadron physics, especially the multihadron system. A typical example is the $\Lambda(1405)$, which is considered as the $\bar{K}N$ quasibound state with strong coupling to the lower $\pi\Sigma$ and $\pi\Lambda$ channels~\cite{Hyodo:2022xhp}. This indicates that the interaction between $\bar{K}$ and $N$ is attractive, making the community interested in the antikaon and multinucleon system. In addition, the interactions between two nucleons for both isospin singlet and triplet channels are also attractive, which makes the antikaon and multinucleon system form a stable multihadron state as discussed in Ref.~\cite{Hyodo:2022xhp}. Along the same line, based on the experimental fact, the hadron physics community focuses on the $DD(DK)$~\cite{Wu:2019vsy,MartinezTorres:2018zbl,Pang:2020pkl}, $\bar{D}\bar{D}^*\Sigma_c$~\cite{Pan:2022xxz,Wu:2021gyn}, $D^{(*)}D^{(*)}D^{(*)}$~\cite{Wu:2021kbu,MartinezTorres:2020hus}, and other similar systems. For detailed discussions, we recommend Ref.~\cite{Liu:2024uxn}. To study such three-body systems, several traditional methods have been employed, for instance, solving the Schr\"odinger equation using the Gaussian expansion method or Born-Oppenheimer approximation, solving the Faddeev equation in fixed center approximation, the QCD sum rule approach, and so on~\cite{Liu:2024uxn}. The above methods are easily implemented to the three-body system. However, when the number of the system is larger than 3 or the direct three-body interaction is considered, the above methods will be unlikely applied. 
For the latter case, the lattice QCD calculation~\cite{Ichie:2002dy, Bissey:2006bz} has indicated that the three-body force in three quark systems is important.

The hadronic molecular picture suggests that the high-energy intrinsic structures of the hadrons can be integrated out, leaving us with an effective field theory (EFT) of low-energy point particles.
There are abundant applications of the EFT for building the nuclear forces~\cite{Epelbaum:2008ga, Machleidt:2011zz, Epelbaum:2019kcf, Machleidt:2024bwl} and the hadron-hadron interactions~\cite{Haidenbauer:2023qhf, Petschauer:2020urh, Song:2018qqm}.
The corresponding few- and many-body Schr\"odinger equations are then solved for nuclear binding energies and other low-energy observables.
In this work, we apply the framework of the nuclear lattice effective field theory (NLEFT), by which we discretize the space with a cubic lattice and build the discrete version of the chiral interactions among the particles~\cite{Lee:2008fa}. 
The resulting lattice Hamiltonian can be solved with various first-principles algorithms.
In this work, we only consider the three-body system for which the exact sparse matrix diagonalization methods such as the Lanczos method are available~\cite{Borasoy:2005yc, Epelbaum:2009zsa}. 
However, for particle numbers greater than 3, we have to consider more advanced quantum Monte Carlo algorithms~\cite{Borasoy:2006qn, Elhatisari:2017eno, Lu:2021tab}.
Contrasted with the shell-model based nuclear {\it ab initio} methods, the NLEFT works directly in the coordinate representation and has shown advantages on phenomena with strong correlations such as nuclear clustering~\cite{Epelbaum:2012qn, Epelbaum:2013paa, Elhatisari:2016owd, Freer:2017gip} and nuclear thermodynamics~\cite{Lu:2019nbg, Ma:2023ahg}. 
\begin{figure}[h]
\centering
    \includegraphics[width=0.48\textwidth]{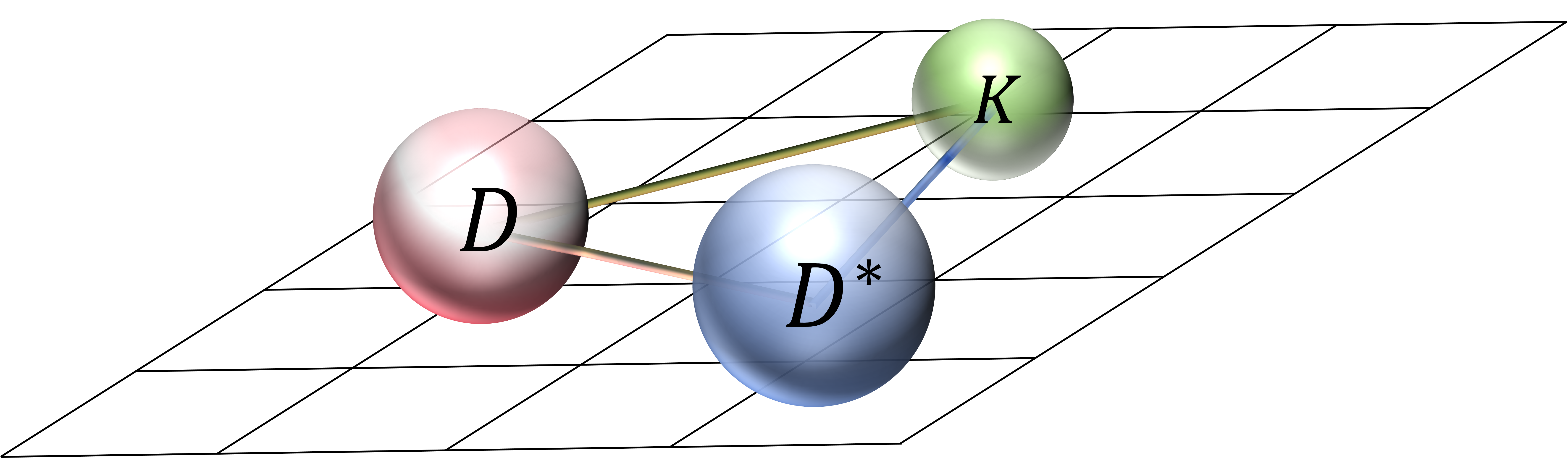}
    \caption{The schematic diagram of the $DD^*K$ system on the lattice. The sizes of the balls signify the corresponding masses.}
    \label{fig:schematic_plot}
\end{figure}

The application of the NLEFT for general hadronic systems [hadronic lattice EFT, (HLEFT)] just began recently.
A novel impurity lattice Monte Carlo method was developed for simulating a few hyperons immersed into a large number of nucleons~\cite{Elhatisari:2014lka, Bour:2014bxa}.
The hypernuclei with one~\cite{Frame:2020mvv, Hildenbrand:2024ypw} or two $\Lambda$ hyperons~\cite{Hildenbrand:2022imw} and neutron matter with hundreds of hyperons~\cite{Tong:2024egi} have been simulated with the HLEFT.
In this work, we extend the calculation to multiple hadrons with unequal masses. 
This scenario immediately introduces two folds of complexities.
First, the two-body interactions are different for various hadron pairs. 
For each pair of hadrons, we have to make a separate determination of the low-energy constants whose number might be large.
To this end, here we limit our calculations to simple leading-order interactions.
Second, usually we lack the required data for parametrizing the interactions. 
For example, if the three-body system we are interested in has not been detected, it is difficult to pin down the strength of the three-body force (3BF), which has played an essential role in nuclear binding~\cite{Hammer:2012id}.
As an alternative, here we only use the three-body force as a diagnosis tool for estimating the systematic uncertainties.
Note that these higher-body forces can be generated by sliding the cutoffs~\cite{Bogner:2009bt}; examining the impact of the 3BF also provides important insights into the renormalization group invariance of the EFT interactions.

As an exploratory study of the HLEFT, this work will solve the $DD^*K$ system (as illustrated in Fig.~\ref{fig:schematic_plot}) which comprises three experimental well-established two-body subsystems, i.e., the $T_{cc}^+$~\cite{LHCb:2021vvq,LHCb:2021auc}, the $D_{s0}^*(2317)$~\cite{Belle:2003guh,CLEO:2003ggt}, and the $D_{s1}(2460)$~\cite{Belle:2003guh,CLEO:2003ggt}.  In this work, we construct the two-body interactions and the three-body interaction with respect to the heavy quark symmetry and chiral symmetry. The parameters of the two-body interactions are determined by the binding energies of the three mentioned subsystems, while the three-body force is kept undetermined. According to this strategy,  the interactions are constructed in Sec.~\ref{sec:Methods}. The details of lattice calculations, results, and discussions are given in Sec.~\ref{sec:results_and_discussions}. The summary is given in Sec.~\ref{sec:Summary}.

\section{Framework}
\label{sec:Methods}

\subsection{The $DD^{\ast}$ two-body interaction}
The peak, named as $T_{cc}^+$, in the $D^0D^0\pi^+$ invariant mass distribution is very close to the $DD^*$ threshold, which indicates it as a $DD^*$ hadronic molecular candidate. Meanwhile, the absence of the peak structure in the $D^+D^+$ invariant mass distribution demonstrates that $T_{cc}^+$ does not have an isospin partner and could be an isospin singlet state. Based on this fact, we work in the isospin singlet scenario with both contact potential and one-pion-exchanged (OPE) potential. 

One can construct leading-order (LO) contact interaction~\cite{Du:2021zzh,Chen:2021vhg,Qiu:2023uno}
\begin{align}
    \mathcal{L}_{DD^{\ast}}=&-\frac{D_{00}}{8}\langle H_a^\dag H_bH_b^\dag H_a\rangle\notag\\
    &-\frac{D_{01}}{8}\langle \sigma^iH_a^\dag H_b\sigma^iH_b^\dag H_a\rangle\notag\\
    &-\frac{D_{10}}{8}\langle \tau^A_{aa'}H_{a'}^\dag H_b\tau^A_{bb'}H_{b'}^\dag H_a\rangle\notag\\
    &-\frac{D_{11}}{8}\langle \tau^A_{aa'}\sigma^iH_{a'}^\dag H_b\tau^A_{bb'}\sigma^iH_{b'}^\dag H_a\rangle,
\label{contact_DDs_lagrangian}
\end{align}
with respect to chiral symmetry and heavy quark spin symmetry.
Here the subscripts $a^{(')}$, $b^{(')}$ denote flavor indices, and $\tau^{A=1,2,3}$
are the isospin Pauli matrices. The coefficients $D_{00,10,01,11}$ are four
low-energy constants (LECs) describing the contact interactions between the ground charmed doublet
\begin{align}
    H_a=P_a+\bm{V}_a\cdot\bm{\sigma},
\end{align}
with $\bm{\sigma}$ the Pauli matrices. Here $P_a$ and $\bm{V}_a$ only contain the annihilating operators of the ground pseudoscalar and vector charmed mesons, respectively. Expanding in flavor space, $P_a$ and $V_a$ can be written explicitly as
\begin{align}
    P_a=&\left(\begin{array}{c}
         D^0  \\
          D^+
    \end{array}\right)_a,\ \ \ 
    \bm{V}_a=\left(\begin{array}{c}
         \bm{D}^{\ast0}  \\
          \bm{D}^{\ast+}
    \end{array}\right)_a.
\end{align}
By constructing isospin singlet $DD^*$ wave function
\begin{align}
    |DD^\ast,I=0\rangle=-\frac{1}{\sqrt{2}}(|D^0D^{\ast+}\rangle-|D^+D^{\ast0}\rangle),
\end{align}
one can obtain its contact potential 
\begin{align}
    V_{\text{Con}}^{I=0}(DD^\ast\to DD^\ast)=&-2(D_{01}-3D_{11})\equiv v_0,
\end{align}
from the Lagrangian Eq.~\eqref{contact_DDs_lagrangian}. It is noted that the potential is a linear combination of the LECs $D_{01}$ and $D_{11}$ and can be redefined as another parameter $v_0$. With this new redefined parameter, the isospin singlet $|DD^*\rangle$ contact interaction reads as 
\begin{align}
    V^{\text{Con}}_{DD^\ast}&=v_0\bm{\epsilon}\cdot\bm{\epsilon}^\ast.
    \label{contact_DDs_potential}
\end{align}
Considering the $T_{cc}^+$ is approaching the $DD^*$ threshold, we deal with the $D^*$ polarization vector $\bm{\epsilon}^{(\ast)}$ non-relativistic, i.e. 
$\bm{\epsilon}_0=(0, 0, 1)$, $\bm{\epsilon}_{\pm 1} = \mp \frac{1}{\sqrt{2}}
(1, \pm i, 0)$ with subindex for its helicity. After the partial wave projection, one can obtain the $S$-wave contact potential $V^{\text{Con}}_{DD^\ast}=v_0$ in momentum space. The corresponding potential in coordinate space is a $\delta$ function $V^{\text{Con}}_{DD^\ast}(\bm{r})=v_0\delta(\bm{r})$.

The LO Lagrangian for the $DD^\ast \pi$ interaction reads as
\begin{align}
    \mathcal{L}_{DD^\ast \pi}=\frac{1}{4}g\langle\bm{\sigma}\cdot\bm{u}_{ab}H_bH_a^\dag\rangle,
\end{align}
where $\bm{u}=-\nabla\bm{\Phi}/f_\pi$ with
\begin{align}
    \bm{\Phi}=\left(\begin{array}{cc}
       \pi^0  & \sqrt{2}\pi^+ \\
        \sqrt{2} \pi^- &-\pi^0
    \end{array}\right),
\end{align}
and the pion decay constant $f_\pi=92.2~\mathrm{MeV}$. Here the coupling $g=0.57$ is extracted from the experimental decay width of the $D^{\ast+}\to D^0\pi^+$ process~\cite{Du:2021zzh}.
The corresponding OPE potential for isospin singlet $|DD^{\ast}\rangle$ in momentum space is ~\cite{Du:2021zzh}
\begin{align}
    V^{OPE}_{DD^\ast}(\bm{q})=-\frac{3g^2}{4f_\pi^2}\frac{\bm{\epsilon}\cdot\bm{q}\ \bm{\epsilon}^{\ast}\cdot\bm{q}}{\bm{q}^{2}+\mu^2}.
    \label{eq:ope_DDs_potential_q}
\end{align}
Here $\mu^2=M_\pi^2-(M_{D^\ast}^2-M_{D}^2)$, $\bm{q}=\bm{p}-\bm{p}'$ is the transferred three-momentum, and $\bm{p}$ and $\bm{p}'$ are the relative incoming and outgoing three-momentum, respectively. The potential in coordinate space 
\begin{align}
    V^{OPE}_{DD^\ast}(\bm{r})=-\frac{3g^2}{4f_\pi^2}(\frac{\bm{\epsilon}\cdot\bm{\epsilon}^{\ast}}{3}\bm{\delta}(\bm{r})-v_C(\bm{r})\bm{\epsilon}\cdot\bm{\epsilon}^{\ast}+v_T(\bm{r})\bm{S}(\bm{r}))
\end{align}
can be obtained by a Fourier transformation. 
Here $v_C(\bm{r})$ and $v_T(\bm{r})$ are the central and tensor potentials, respectively,  which read
\begin{align}
   v_C(\bm{r})&\equiv\frac{\mu^3}{12\pi}\frac{e^{-\mu r}}{\mu r},\\
   v_T(\bm{r})&\equiv v_C(r)(1+\frac{3}{\mu r}+\frac{3}{\mu^2r^2}). 
\end{align}
 The tensor operator $\bm{S}(\bm{r})$ is defined as 
\begin{align}
   \bm{S}(\bm{r})\equiv \frac{3(\bm{\epsilon}\cdot\bm{r})(\bm{\epsilon}^{\ast}\cdot\bm{r})}{r^2}-\bm{\epsilon}\cdot\bm{\epsilon}^{\ast}.
\end{align} 
After the partial wave projection, one can obtain the S-wave OPE potential\footnote{When $M_\pi< M_{D^*}-M_D$, the denominator of Eq.~\eqref{eq:ope_DDs_potential_q} can equal zero with an appropriate positive $|\mathbf{q}|$ value, which brings an imaginary part to the OPE potential. We have checked that this imaginary part has small impact on the binding energy, especially on the real part of the binding energy. As a result, we only consider the real part of the OPE potential of 
 Eq.~\eqref{eq:ope_DDs_potential_r}.}
\begin{align}
    V^{\text{OPE}}_{DD^\ast}(\bm{r})=\frac{3g^2}{4f_\pi^2}(v_C(\bm{r})-\frac{\bm{\delta}(\bm{r})}{3}).
    \label{eq:ope_DDs_potential_r}
\end{align}
In total, the potential in coordinate space reads as 
\begin{align}
    V_{DD^\ast}(r)=V^{\text{Con}}_{DD^\ast}(r)+V^{\text{OPE}}_{DD^\ast}(r),
    \label{eq:DDs_potential}
\end{align}
with only one free parameter $v_0$ embodied in the contact potential.

\subsection{The $DK$ and $D^{\ast}K$ two-body interactions}

The LO $DK$ and $D^{\ast}K$ interactions can be extracted by introducing covariant derivative 
\begin{align}    
\mathcal{D}_{\mu}&=\partial_\mu+\Gamma_\mu,
\end{align}
to the kinetic and the mass terms of the charmed mesons~\cite{Guo:2009ct}
\begin{align}
    \mathcal{L}_{\text{LO}}=&\mathcal{D}_{\mu}D\mathcal{D}^{\mu}D^\dag-\mathring{M}_D^2DD^\dag.
\label{lo_DK_lagrangian}
\end{align} 
Here $D=(D^0,D^+,D^+_s)$, and $\mathring{M}_D$ is charmed mesons
with the chiral limit mass. The building block $\Gamma_\mu$ is defined as 
\begin{align}
    \Gamma_\mu&\equiv\frac{1}{2}(u^\dag\partial_\mu u+u\partial_\mu u^\dag),
\end{align}
with 
\begin{align}
    u=\exp(\frac{i\phi}{\sqrt{2}f_{\pi}}).
\end{align}
The pseudoscalar octet are collected in a matrix form
\begin{align}
    \phi=\left(\begin{array}{ccc}
        \frac{1}{\sqrt{2}}\pi^0+\frac{1}{\sqrt{6}}\eta &  \pi^+&K^+\\
        \pi^- & -\frac{1}{\sqrt{2}}\pi^0+\frac{1}{\sqrt{6}}\eta &K^0\\
        K^-&K^0&-\frac{2}{\sqrt{6}}\eta
    \end{array}\right).
\end{align}

The $D^\ast_{s0}(2317)$ and $D_{s1}(2460)$ are with the same distance to the $DK$ and $D^\ast K$ thresholds, respectively, which indicates that they could be the corresponding hadronic molecules~\cite{Guo:2009ct,vanBeveren:2003kd,MartinezTorres:2011pr}, and the results for lattice QCD also support the above model~\cite{Liu:2012zya}. Their isoscalar properties make us focus on the $I=0$ $DK$ and $D^\ast K$ interactions. Based on the Lagrangian discussed above, one can extract the S-wave isoscalar $DK$ potential  
\begin{align}
    V_{\text{LO}}^{DK}(p_i)= \frac{-1}{2f_\pi^2}(p_1\cdot p_2+p'_1\cdot p'_2+p_1\cdot p'_2+p_2\cdot p'_1),
    \label{eq:LO_DK_potential}
\end{align}
with $p_{1,2}$ and $p'_{1,2}$ the four momenta of the incoming and outgoing particles, respectively. The small binding energy of the $D^\ast_{s0}(2317)$ allows us to make a non-relativistic approximation $p_i^0= m_i+\frac{\bm{p}^2_i}{2m_i}$ for the energy $p_i^0$ of the $i$th particle to simplify the potential on the lattice. 

As the binding energy of the $D^\ast_{s0}(2317)$ is larger than that of the $T_{cc}^+$, we need to further consider the next-leading-order (NLO) chiral Lagrangian for the interaction between the pseudoscalar charmed mesons and the Goldstone bosons
\begin{align}
    \mathcal{L}_{\text{NLO}}&=D(-h_0\langle\chi_+\rangle-h_1\tilde{\chi}_++h_2\langle u_\mu u^\mu\rangle-h_3u_\mu u^\mu)\bar{D}\notag\\
    &+\mathcal{D}_{\mu}D(h_4\langle u^\mu u^\nu\rangle-h_5\{u^\mu, u^\nu\}-h_6[u^\mu, u^\nu])\mathcal{D}_{\nu}\bar{D},
\label{nlo_DK_lagrangian}
\end{align}
where
\begin{align}
    \chi_+&=u^\dag\chi u^\dag+u\chi u,\notag\\
    \tilde{\chi}_+&=\chi_+-\frac{1}{3}\langle\chi_+\rangle,\notag\\
    u_\mu&=iu^\dag\mathcal{D}_{\mu}Uu^\dag.
\end{align}
The quark mass matrix is diagonal $\chi=2B\cdot\mathrm{diag}(m_u,m_d,m_s)$ with $B=|\langle0|q\bar{q}|0\rangle|/f_\pi^2$ and $\langle0|q\bar{q}|0\rangle$ the quark condensate. $h_i(i=0,\cdots,6)$ are unknown coefficients. In the following, we drop the $h_0$, $h_2$, and $h_4$ terms, which are suppressed in the large-$N_c$ limit of QCD as discussed in Ref.~\cite{Lutz:2007sk}. The $h_6$ term can also be neglected since it is suppressed by one order due to the commutator structure~\cite{Guo:2009ct}. 
In this case, the $I=0$ $DK$ NLO contact potential in S-wave can be written as~\cite{Guo:2009ct}
\begin{align}
    V_{\text{NLO}}^{DK}(p_i)= &-\frac{8M_{K}^2}{3f_\pi^2}h_1+\frac{4}{f_\pi^2}(h_3p_2\cdot p'_2\notag\\
    &+h_5(p_1\cdot p_2p'_1\cdot p'_2+p_1\cdot p'_2p_2\cdot p'_1)).
\end{align}
Here $M_{K}$ is the kaon mass. The coefficient $h_1=0.42$ can be determined from the mass differences among the $D$ mesons. As the coefficients $h_3$ and $h_5$ have a strong correlation, one can fix one of them. In our case, we fix $h'_5=1$ and fit $h_3$ as that in Ref.~\cite{Guo:2009ct}. Here we redefine $h'_5\equiv h_5M^2_{D}$. After performing a non-relativistic expansion, one can obtain a simple potential.
In total, the S-wave $DK$ potential is 
\begin{align}
    V_{DK}(\bm{p})=\frac{1}{4M_DM_{K}}
    (V_{\text{LO}}^{DK}(\bm{p})+V_{\text{NLO}}^{DK}(\bm{p})),
    \label{eq:DKb_potential}
\end{align}
with one free parameter $h_3$. The factor $1/(4M_DM_{K})$ is used to match the dimension.

The S-wave $D^\ast K$ LO interaction has the same form as the S-wave $DK$ LO interaction, $i.e.$, $V_{\text{LO}}^{D^\ast K}=V_{\text{LO}}^{DK}$, but different value of the parameters, which we define as $h_1^\ast$, $h_3^\ast$, and $h_5^\ast$. The NLO $D^\ast K$ interaction reads as 
\begin{align}
    V_{\text{NLO}}^{D^\ast K}(p_i)= &(-\frac{8M_K^2}{3f_\pi^2}h^\ast_1+\frac{4}{f_\pi^2}(h^\ast_3p_2\cdot p'_2\notag\\
    &+h^\ast_5(p_1\cdot p_2p'_1\cdot p'_2
    +p_1\cdot p'_2p_2\cdot p'_1)))\bm{\epsilon}\cdot\bm{\epsilon}^{\ast}.
\end{align}
Here we set $h^\ast_{1}=h_1=0.42$, $h'^\ast_5=h^\ast_5M^2_{D^\ast}=1$. Eventually, the $D^\ast K$ potential $V_{D^\ast K}$ reads as
\begin{align}
    V_{D^\ast K}(\bm{p})=\frac{1}{4M_{D^\ast}M_{K}}(V_{\text{LO}}^{D^\ast K}(\bm{p})+V_{\text{NLO}}^{D^\ast K}(\bm{p})),
    \label{eq:DsKb_potential}
\end{align}
by only one free parameter $h^\ast_3$. One can see that the $DK$ and $D^\ast K$  potentials are more complicated than that of the $DD^\ast$ system. We perform a fast Fourier transformation numerically to obtain the corresponding potential in coordinate space.

\subsection{The $DD^{\ast}K$ three-body interaction}

The $DD^\ast K$ three-body interaction can be constructed by the famous axial vector current~\cite{Casalbuoni:1996pg} 
\begin{eqnarray}
\mathcal{A}_\mu=\frac{1}{2}\left(u^\dagger \partial_\mu u-u \partial_\mu u^\dagger\right)\simeq \frac{i}{\sqrt{2}f_\pi}\partial_\mu \phi,
\end{eqnarray}
with $u=\exp\left(\frac{i\phi}{\sqrt{2}f_\pi}\right)$. Its property under chiral transformation $U$ reads as
\begin{eqnarray}
    \mathcal{A}_\mu\to U\mathcal{A}_\mu U^\dagger.
\end{eqnarray}
The superfield of the S-wave ground charmed meson doublet behaves as
\begin{eqnarray}
    H_a\to S H_a,\quad H_a^\prime \to -H_a,\quad H_a\to H_b U_{ba}^\dagger,
\end{eqnarray}
under the heavy quark spin $S$, parity $P$ and chiral $U$ transformations. 
The covariant derivative transfer as
\begin{eqnarray}
    \mathcal{D}_\mu \to U\mathcal{D}_\mu U^\dagger,
\end{eqnarray}
under chiral transformation $U$. 
Here, we redefine the covariant derivative $\mathcal{D}_\mu=\overset{\leftrightarrow}{\partial}_\mu+\Gamma_\mu$, $\overset{\leftrightarrow}{\partial}_\mu=\overset{\rightarrow}{\partial}_\mu-\overset{\leftarrow}{\partial}_\mu$. As the result, the leading-order three-body Lagrangian reads as
\begin{eqnarray}
    \mathcal{L}=c_3\langle H\mathcal{D}_\mu H^\dagger H \mathcal{D}^\mu H^\dagger \rangle+c_3^\prime \langle H\mathcal{A}_\mu H^\dagger H \mathcal{A}^\mu H^\dagger \rangle.
\end{eqnarray}
The second term gives two kaon momenta. The first term gives one kaon momentum and one charmed meson momentum, making its contribution the leading-order contribution. As the result, we consider the first term as the LO three-body Lagrangian, i.e.,
\begin{eqnarray}
    \mathcal{L}_{DD^\ast K}=c_3\langle H\mathcal{D}_\mu H^\dagger H \mathcal{D}^\mu H^\dagger \rangle,
\end{eqnarray}
by one free parameter $c_3$. Accordingly, the LO three-body potential is 
\begin{align}
    V_{DD^\ast K}(p_i)=&\frac{c_3}{4f_\pi^2}(p_1\cdot p_3+p_1\cdot p'_3+p_2\cdot p_3+p_2\cdot p'_3\notag\\
    &+p'_1\cdot p_3+p'_1\cdot p'_3+p'_2\cdot p_3+p'_2\cdot p'_3)\bm{\epsilon}\cdot\bm{\epsilon}^{\ast}.
    \label{eq:DDsKb_potential}
\end{align}
The momentum is also dealt in non-relativistic approximation analogous to that of the $D^{(*)}K$ interaction.

\section{Results and discussions}
\label{sec:results_and_discussions}

To calculate the $DD^\ast K$ system on the lattice, one needs to implement the potentials among them, either in coordinate space or in momentum space. 
The two-body potentials of the $DD^\ast$, $DK$ and $D^\ast K$ systems can be found in Eqs.~\eqref{eq:DDs_potential}, \eqref{eq:DKb_potential}, and \eqref{eq:DsKb_potential}, respectively, with the free parameters $v_0$, $h_3$ and $h_3^*$ fitted by their two-body binding energies. The Hamiltonians describing their two-body interactions are  
\begin{align}
    H_{T_{cc}^+}=&M_{D}+M_{D^\ast}+K_{DD^\ast}+f_{2B}(\bm{p}_i,\bm{p}'_i)V_{DD^\ast}^{\text{Con}}+V_{DD^\ast}^{\text{OPE}},\notag\\
    H_{D^\ast_{s0}}=&M_{D}+M_{K}+K_{DK}+f_{2B}(\bm{p}_i,\bm{p}'_i)V_{DK},\notag\\
    H_{D_{s1}}=&M_{D^\ast}+M_{K}+K_{D^\ast K}+f_{2B}(\bm{p}_i,\bm{p}'_i)V_{D^\ast K},
    \label{eq:Hamiltonian_two_body}
\end{align}
where $K=\sum_{i=1}^2\bm{p}_i^2/2m_i$ is the kinetic energy in the center-of-mass frame. 
Here $\bm{p}_i$ and $m_i$ are the three-momentum and mass of the $i$th particle in the two-body system. For each short-range potential, a single-particle regulator~\cite{Lu:2023jyz}  $f_{2B}(\bm{p}_i,\bm{p}'_i)=\prod_{i=1}^2g_\Lambda(\bm{p}_i)g_\Lambda(\bm{p}'_i)$ is applied,  
 where $g_\Lambda(\bm{p})=\exp(-\bm{p}^6/2\Lambda^6)$ is a soft cutoff function with $\bm{p}_i$ and $\bm{p}'_i$ the incoming and outgoing momenta of the individual particle. Here the long-range part, i.e., the first term of Eq.~\eqref{eq:ope_DDs_potential_r}, of the OPE potential is kept in our calculation, as the short-range part can be absorbed into the definition of the contact potential $V_{DD^\ast}^{\text{Con}}$. 
\begin{figure}[h]
\centering
    \includegraphics[width=0.45\textwidth]{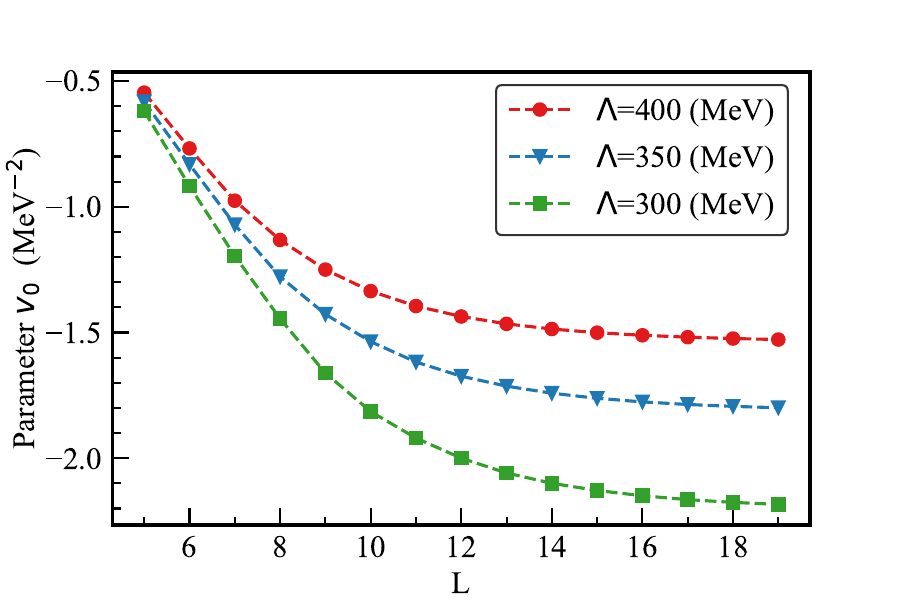}
    \includegraphics[width=0.45\textwidth]{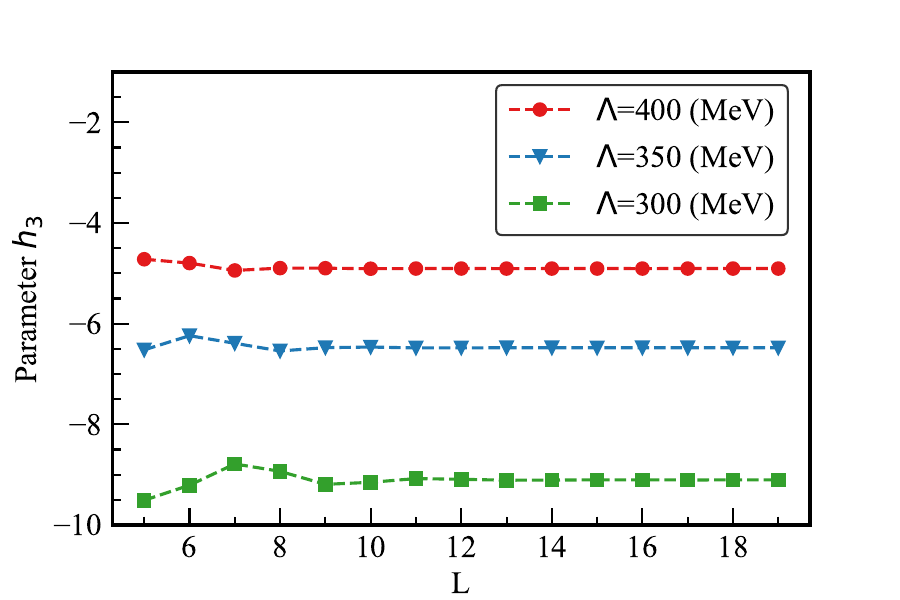}
    \includegraphics[width=0.45\textwidth]{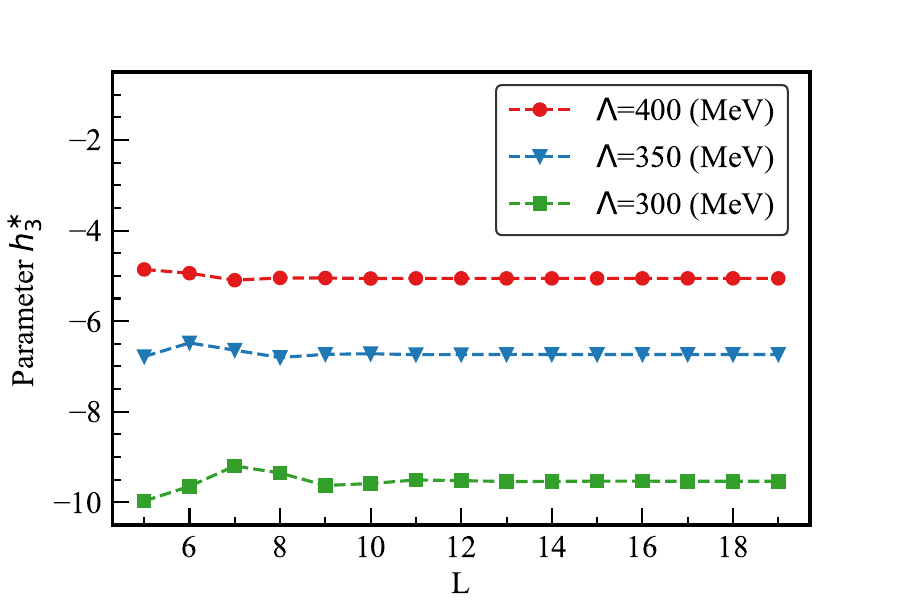}
    \caption{The dependence of the two-body interaction parameters, i.e., $v_0$, $h_3$, $h_3^*$ for the $DD^*$, $DK$, $D^*K$ interactions respectively, on the lattice volume $L$. The red circles, blue inverted triangles, green squares are for the lattice cutoff parameters $\Lambda=400$, $350$, $300~\mathrm{MeV}$, respectively. The values of these parameters are stable with sufficient large volume $L$.}
    \label{fig:two_body_force}
\end{figure}

 With the two-body potentials discussed above, we can perform simulations of these two-body systems.
 In more detail, we use the effective field theory Hamiltonian of Eq.~\eqref{eq:Hamiltonian_two_body} and transform it from momentum space to coordinate space by Fourier transforms. The solution of the Schr$\ddot{\mathrm{o}}$dinger equation can be written explicitly in the single-particle basis $|\phi\rangle=|c_1,\cdots,c_N\rangle$~\cite{Lu:2019nbg}, where $c_i=(\bm{n}_i,\sigma_i)$ are the quantum numbers of the $i$th particle. Here  $\bm{n}_i$ is an integer triplet specifying the lattice coordinates and $\sigma_i$ is the spin. The matrix exact diagonalization scheme uses the implicitly restarted Lanczos method to find the eigenvalues and eigenvectors~\cite{lehoucq1998} by SciPy~\cite{Virtanen:2019joe} in Python.
We perform simulations on $L^3=5^3,6^3,\cdots,19^3$ cubic lattice with $N=2$ bosons and the spatial lattice spacing is chosen as $a=1/200\ \mathrm{MeV}^{-1}\approx0.99~\mathrm{fm}$. The lattice spacing $a$ should not be set smaller than the typical hadron size, as hadrons are degrees of freedom in lattice EFT. The three values of the soft cutoff $\Lambda=400,~350,~300~\mathrm{MeV}$ are chosen to check the renormalization invariance. From the EFT point of view, one cannot set the limit $\Lambda\to\infty$ as new physics may emerge at certain hard scales where the EFT breaks down. The binding energies $E_b= -0.98,~-45.09$ and $-44.70$ MeV of the $T_{cc}^+$, $D^\ast_{s0}(2317)$, and $D_{s1}(2460)$ are used to fix the free parameters $v_0$, $h_3$ and $h_3^*$, respectively, for a given cubic lattice $L^3$. Here the spin averaged mass of the $D$, $D^*$, $K$ mesons are used~\cite{ParticleDataGroup:2008zun}.
More details can be found in Appendix A. The values of these three parameters are plotted in Fig.~\ref{fig:two_body_force} as a function of $L$ for the three values of $\Lambda$. Each point represents a separate simulation. From the figure, one can see that with increasing $L$ all three parameters converge gradually in $\Lambda=300,350,400~\mathrm{MeV}$, but with different trends. The parameters of $\Lambda=400~\mathrm{MeV}$ converge faster than those of the other two $\Lambda$ values. That is because that the interaction range is set by $1/\Lambda$. A larger $\Lambda$ value corresponds to a smaller interaction range, making the results converge more quickly on the lattice. The dependence of the dimensionless parameters $h_3^{(*)}$ on the volume size $L^3$ is different from that of the parameter $v_0$. That is because the other parameters, i.e., $h_1^{(*)}$,  $h_5^{(*)}$, etc., are fixed to their empirical values for the $D^{(*)}K$ systems and the $DD^*$ system only has one free parameter $v_0$. As the uncertainty from the exact diagonalization scheme is very small, we only present the systematic uncertainty from the finite volume size.   

\begin{figure}[h]
\centering
    \includegraphics[width=0.48\textwidth]{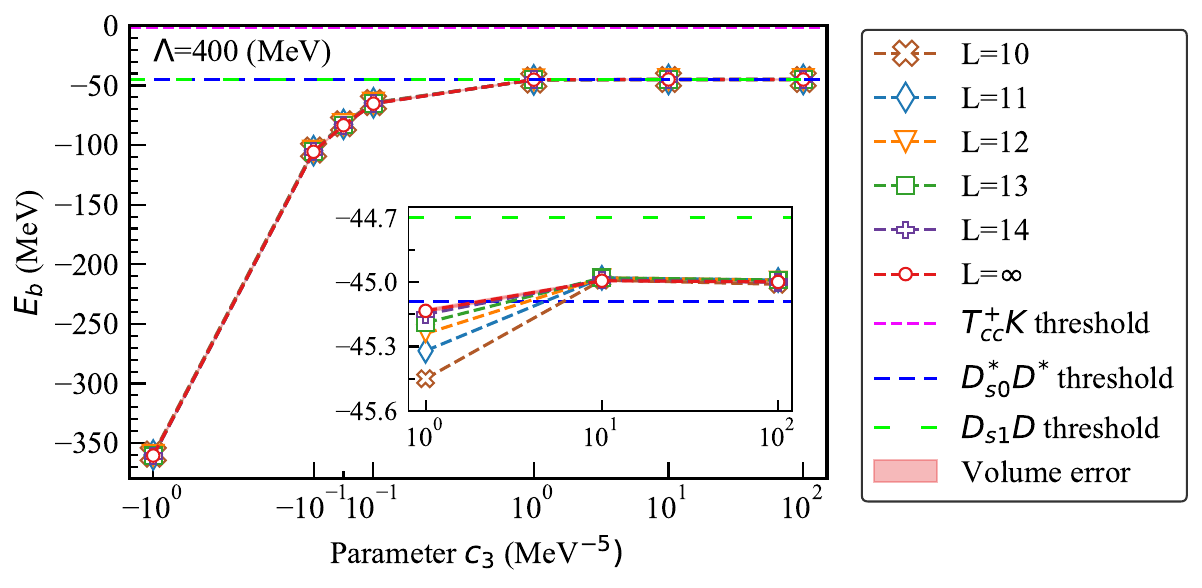}
    \includegraphics[width=0.48\textwidth]{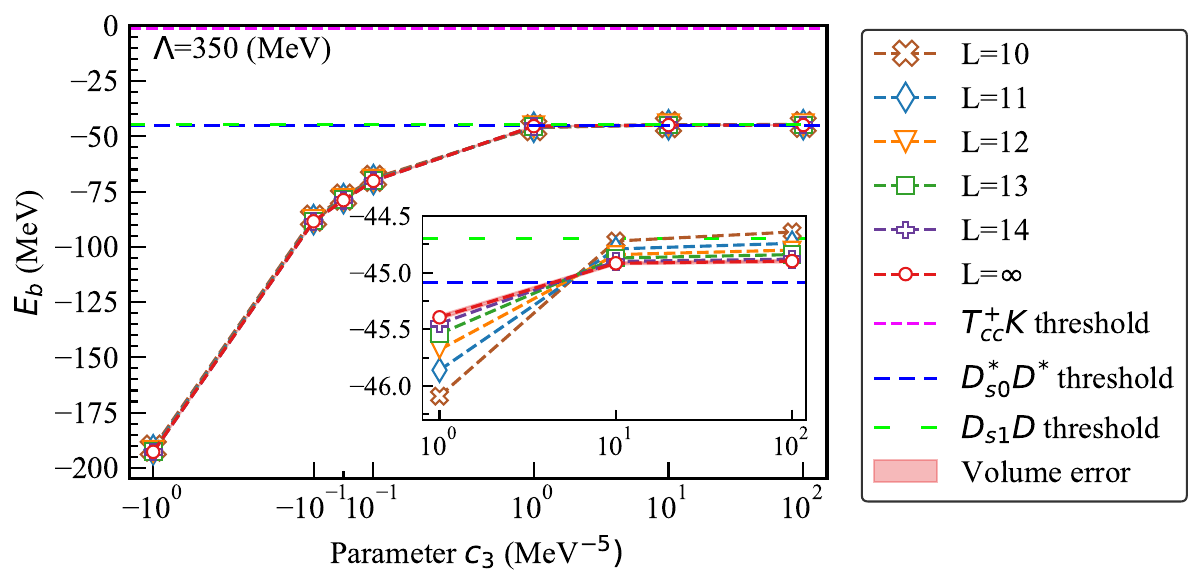}
    \includegraphics[width=0.48\textwidth]{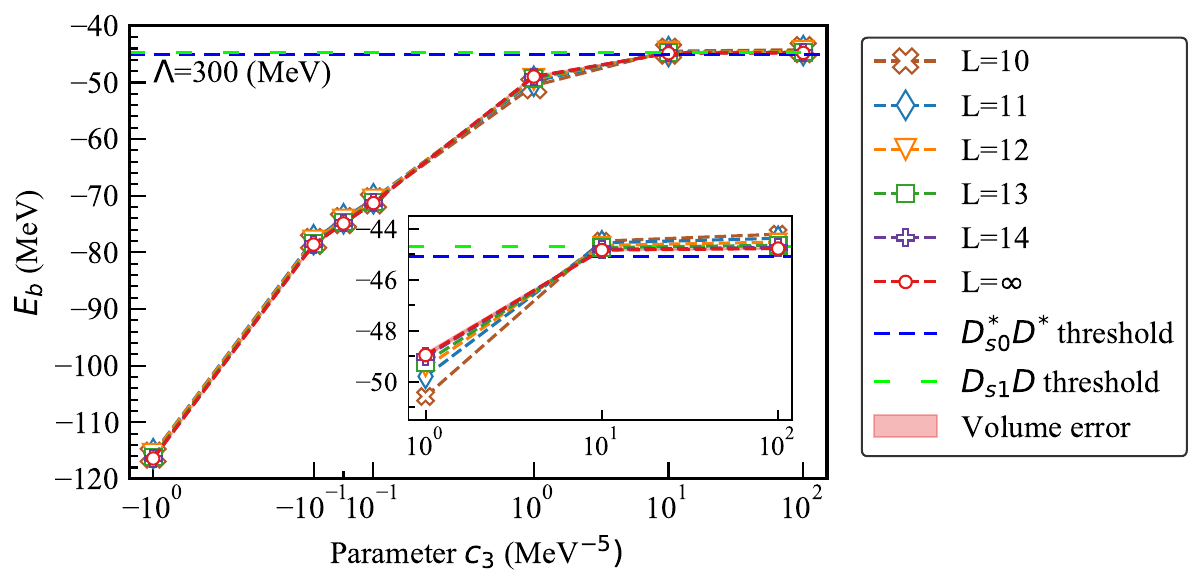}
    \caption{The three-body binding energy $E_b$ on the three-body interaction parameter $c_3$. The orange crosses, blue diamonds, yellow inverted triangles, green squares, and purple plus signs are for volume $L=10,11,12,13,14$ in order. The red circles and bounds are the central value and the uncertainty when $L=\infty$ by using the extrapolation formula Eq.~\eqref{eq:volume_effect}. The horizontal pink short dashed, blue dashed, green long dashed lines are for the two-body thresholds $T_{cc}^+K$, $D_{s0}^*D^*$, $D_{s1}D$ thresholds, respectively. The upper, middle and lower panels are for lattice cutoff parameters $\Lambda=400$, $350$, $300~\mathrm{MeV}$, respectively.}
    \label{fig:three_body}
\end{figure}

With the three two-body interaction parameters, i.e. $v_0$, $h_3$ and $h_3^\ast$, fitted by the corresponding two-body binding energies, we can proceed to calculate the binding energy of the $DD^\ast K$ three-boson system. The  Hamiltonian is written as
\begin{align}
    H=&M_{D}+M_{D^\ast}+M_{K}+K_{DD^\ast K}+f_{2B}(\bm{p}_i,\bm{p}'_i)V^{\text{Con}}_{DD^\ast}\notag\\
    &+V^{\text{OPE}}_{DD^\ast}+f_{2B}(\bm{p}_j,\bm{p}'_j)V_{DK}+f_{2B}(\bm{p}_k,\bm{p}'_k)V_{D^\ast K}\notag\\
    &+f_{3B}(\bm{p}_l,\bm{p}'_l)V_{DD^\ast K}, 
    \label{eq:Hamiltonian_three_body}
\end{align}
where $K_{DD^*K}=\sum_{i=1}^3\bm{p}_i^2/2m_i$ is the kinematic energy in the $K$ rest frame. Analogous to that for the two-body system, a non-local single particle regulator $f_{3B}(\bm{p}_i,\bm{p}'_i)=\prod_{i=1}^3g_\Lambda(\bm{p}_i)g_\Lambda(\bm{p}'_i)$ is also introduced for short-range three-body interaction. For the three-body system, one can only obtain a reasonable binding energy for much smaller boxes compared with the two-body cases due to the hardware limit. Thus, 
we perform simulations on $L^3=10^3,11^3,\cdots,14^3$ cubic lattice for the three-hadron system. The lattice spacing is set to $a=1/200\ \mathrm{MeV}^{-1}\approx0.99~\mathrm{fm}$ and the cutoff $\Lambda$ is chosen as $400$, $350$, $300~\mathrm{MeV}$, consistent with that of the two-body system. After transforming the Hamiltonian of Eq.~\eqref{eq:Hamiltonian_three_body} from momentum space to coordinate space by the fast Fourier transforms, the matrix exact diagonalization scheme can be applied to extract the binding energy $E_b$ of the $DD^\ast K$ three-body system. 

\begin{table*}
    \centering
    \renewcommand{\arraystretch}{1.5}
    \caption{Mass and binding energies (in units of MeV) of the $DD^\ast K$ bound states (disregard three-body interaction), in comparison with the results of other works.}
    \begin{tabular}{cp{3.2cm}<{\centering}p{3.2cm}<{\centering}p{3.2cm}<{\centering}p{3.cm}<{\centering}p{2.2cm}<{\centering}}
    \hline
    $I(J^P)$ &This work& Ref.~\cite{Wu:2020job}&Ref.~\cite{Ma:2017ery}&Ref.~\cite{Ren:2018pcd}&Ref.~\cite{Ren:2024mjh} \\
    \hline
    $\frac{1}{2}(1^-)$ $DD^\ast K$ & $4292.39^{+4.36}_{-4.16}(79.06^{+4.36}_{-4.16})$&$\cdots$&$4317.92^{+3.66}_{-4.32}(53.52^{+3.66}_{-4.32})$ &$\cdots$&4309(62)\\
    \hline
    $\frac{1}{2}(1^-)$ $D\bar{D}^\ast K$ & $\cdots$&$4294.1^{+3.1}_{-6.6}(77.3^{+3.1}_{-6.6})$&$4317.92^{+6.55}_{-6.13}(53.52^{+6.55}_{-6.13})$ &$4307\pm2(64\pm2)$&$\cdots$\\
    \hline
    \end{tabular}  \label{tab:Redults}
\end{table*}

Currently, there is no experimental data to fix the 
three-body interaction parameter $c_3$.
 Alternatively, one can study the three-body binding energy $E_b$ with respect to this parameter as shown in Fig.~\ref{fig:three_body}. In the calculation, the values $c_3=-1, -0.1, 0.1, 1, 10, 100~\mathrm{MeV}^{-5}$ are used for an illustration. The upper, middle, and lower panels are for $\Lambda=400, 350, 300~\mathrm{MeV}$ in order on the $L^3$ cubic lattice with $L=10, 11, \dots 14$. Each point represents a separate simulation. The negative $c_3$ value means an attractive three-body interaction, leaving the three-body binding energy increasing dramatically with the increasing attractive interaction. This observation can be related to the famous phenomena of Thomas collapse~\cite{Thomas:1935zz}, which states that the three-body system with strong zero-range pairwise interactions is always unstable. As the range of the interactions decreases, the ground state energy is not bounded from below and falls to negative infinity quickly. Our finding is also consistent with Ref.~\cite{Nielsen:2001hbm}, where it was demonstrated that the Thomas effect can occur when at least two of the three pairwise potentials are of zero range. On the other hand, the three-body system can be stabilized by introducing a repulsive three-body force or a positive $c_3$. We found that even when the $c_3$ goes to a positive infinity, the three-body system can still be bounded with sizable binding energy. This behavior is a consequence of the short-range nature of the three-body force used here. As we do not consider more complicated long-range three-body forces involving pion exchanges, here the induced three-body repulsion is local and only affects particles on the same lattice site. Thus with extremely large three-body repulsive forces, every two hadrons can exhibit effective long-range two-body interactions by exchanging the third spectator particle~\cite{Hammer:2010kp, Naidon:2016dpf}. In this regard the observed pattern is generic in three-body systems bounded by short-range interactions.

The computing time 
increases dramatically with large volumes, which goes even beyond the capability of the current computing resource. To eliminate the finite volume effects, one can alternatively extrapolate the results calculated at finite volume to the infinite volume limit. In this work, we calculate the results at small box sizes 
$10\leq L \leq 14$ and extrapolate the results by using the analytical results for the finite volume effect correction of three-body bound states for nonidentical particles~\cite{Meissner:2014dea,Meng:2017jgx},
\begin{align}
    \frac{\Delta E}{E_T}=&-(\kappa L)^{-3/2}\sum_{i=1}^{3}C_i\exp(-\mu_i\kappa L). 
    \label{eq:volume_effect}
\end{align}
Here $E_T$ and $E_L$ are the binding energies at infinite volume and cubic volume size $L^3$, respectively. 
$\Delta E=E_T-E_L$ is their energy difference. 
The binding momentum $\kappa$ is defined as $\kappa=\sqrt{-2ME_T}$ with 
$M=(m_1+m_2+m_3)/6$ the normalization mass. Here $\mu_i$ and $C_i$ are the reduced mass and free parameters for the $i$-th two-body system. More details can be found in Appendix \ref{app:B}.

\begin{figure}[h]
\centering
    \includegraphics[width=0.48\textwidth]{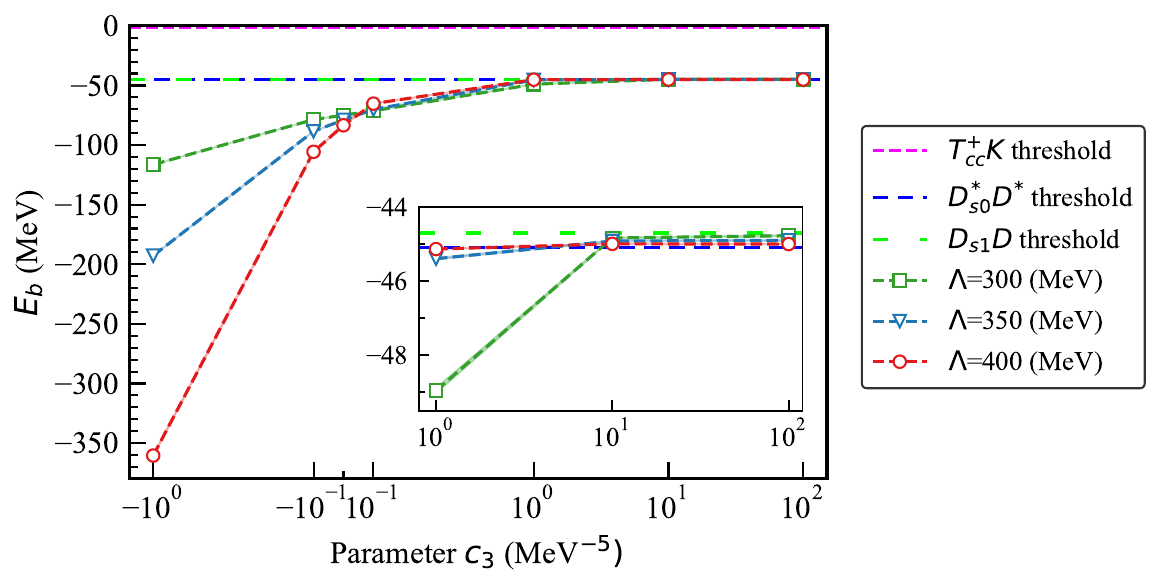}
    \caption{The dependence of the three-body binding energy $E_b$ at infinity volume on the three-body interaction parameter $c_3$. The green squares, blue inverted triangles, red circles are for the cases with lattice cutoff parameter $\Lambda=300$,  $350$, $400~\mathrm{MeV}$, respectively. The corresponding bands are the errors from the fit with the extrapolation formula Eq.~\eqref{eq:volume_effect}. The horizontal pink short dashed, blue dashed, green long dashed lines are for the two-body thresholds $T_{cc}^+K$, $D_{s0}^*D^*$, $D_{s1}D$ in order.}    \label{fig:three_body_lambda}
\end{figure}

The extrapolated binding energy $E_T$'s for various $c_3$'s and $\Lambda$'s are plotted in Fig.~\ref{fig:three_body} with the red circles, with the uncertainties (red bounds) originated from the fitting procedure. One can see that the binding energies are arranged in order of the volume parameter $L$. In this case, one can extract stable binding energy at infinite volume as shown by the red circles in Fig.~\ref{fig:three_body}. To make a comparison for various $\Lambda$'s, we plot the extracted binding energies at infinite volume in Fig.~\ref{fig:three_body_lambda}. 
One can see that no matter whether the three-body interaction is attractive or repulsive, there exist bound states in the $DD^\ast K$ system, even though the exact binding energies are always dependent on the parameter $c_3$. 

The existence of the loosely bound states for the three two-body subsystems, i.e., the $T_{cc}^+$, $D_{s0}^*(2317)$, and $D_{s1}(2460)$, indicates the existence of the three-body bound state, although the exact value of the three-body binding energy depends on the three-body interaction. 
Analogous three-body systems have been well understood in the two limits, i.e, the exact unitary limit (Efimov effect~\cite{Efimov:1970zz}) and the zero-range two-body interaction limit (Thomas collapse~\cite{Thomas:1935zz}). 
The Thomas collapse has also been discussed for the three-nucleon system ~\cite{Yang:2022esu} with the cutoff in the Gaussian regulator to infinity, which means the two nucleons come infinitely close to each other. This collapse can be avoided by including a repulsive three-body interaction at LO~\cite{Bedaque:1998kg}.  

Therefore, in our case, to obtain a stable three-body bound state, a suitable range of $c_3$ is $(0,\infty)~\mathrm{MeV}^{-5}$, which corresponds to a binding energy in the $(-84,-44)~\mathrm{MeV}$ region. In addition, there are model calculations on the $DD^*K$ system without the three-body force $c_3$~\cite{Wu:2020job}. We can also set $c_3$ to zero for a comparison as shown in Tab.~\ref{tab:Redults}. Our calculation gives a binding energy in the infinite volume of $79.06^{+4.36}_{-4.16}~\mathrm{MeV}$, which is in agreement with that of Ref.~\cite{Wu:2020job}. The central value is the average value of those with $\Lambda=400,~350,~300~\mathrm{MeV}$. The upper and lower limit corresponds to the differences between the central value and those with $\Lambda=400,~300~\mathrm{MeV}$.
\begin{table*}
\renewcommand{\arraystretch}{1.5}
    \centering
    \caption{The values of the parameter $c_3$ on various $L^3$ cubic lattice with $\Lambda=400,~350,~300~\mathrm{MeV}$. The binding energies with $\Lambda=400~\mathrm{MeV}$ are inputs. Those with $\Lambda= 350,~300~\mathrm{MeV}$ are outputs.}
    \begin{tabular}{c|c|p{2cm}<{\centering}p{2cm}<{\centering}p{2cm}<{\centering}p{2cm}<{\centering}p{2cm}<{\centering}p{2cm}<{\centering}|p{1cm}<{\centering}}
    \hline
    \multirow{2}*{$\Lambda$ (MeV)}&\multirow{2}*{Parameter}&\multicolumn{6}{c|}{$L$}&\multirow{2}{*}{State}\\
    \cline{3-8}
     &&9&10&11&12&13&14 \\
     \hline
     400&\multirow{3}*{$c_3$ (MeV$^{-5}$)}
         & 0.100& 0.100& 0.100& 0.100& 0.100& 0.100&Input\\
     350&& 0.170& 0.162& 0.164& 0.164& 0.163& 0.163&Fitted\\
     300&& 0.328& 0.305& 0.281& 0.278& 0.281& 0.280&Fitted\\
    \hline
    \end{tabular}  \label{tab:Fit_c3}
\end{table*}

\begin{figure}[h]
\centering
    \includegraphics[width=0.48\textwidth]{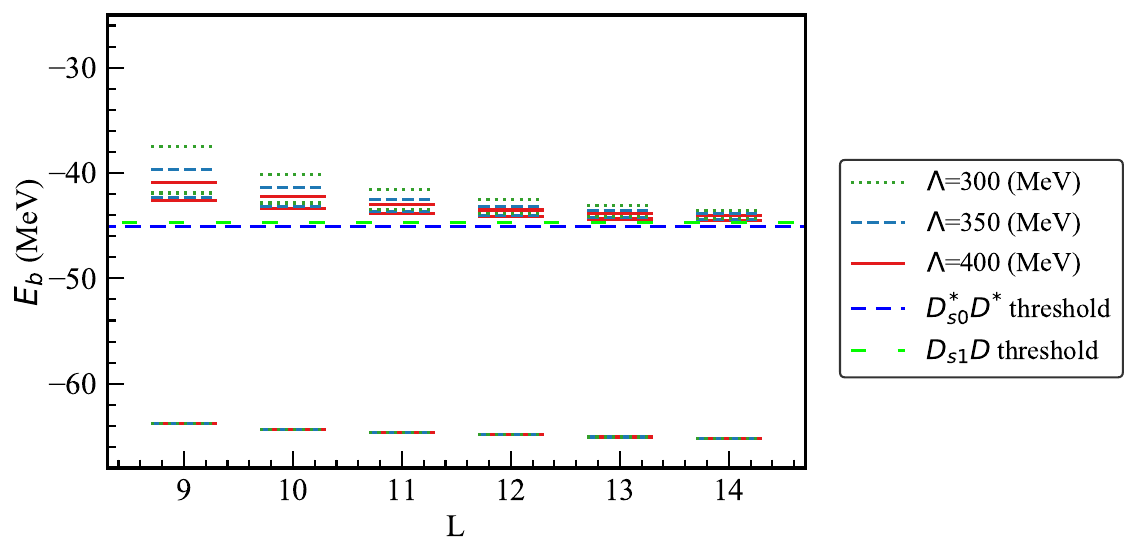}
    \caption{The three-body binding energies of the ground and the first excited state with the cutoff $\Lambda=300~\mathrm{MeV}$ (green dotted curves), $350~\mathrm{MeV}$ (light blue dashed ones), $400~\mathrm{MeV}$ (red solid ones) in terms of the volume $L$.  The horizontal blue dashed, green long dashed lines are for the $D_{s0}^*D^*$, $D_{s1}D$ two-body thresholds, respectively.}
    \label{fig:Efimov_state}
\end{figure}

Without experimental data on the three-body system, the three-body interaction parameter $c_3$ cannot be determined. We use the binding energy with the parameter $c_3=0.1~\mathrm{MeV}^{-5}$ at $\Lambda=400~\mathrm{MeV}$ as an input to check the renormalization invariance of our framework. We fit 
the parameter $c_3$s (shown in Tab.~\ref{tab:Fit_c3}) at $\Lambda=350$ and $\Lambda=300~\mathrm{MeV}$ to obtain the same binding energy. Matrix exact diagonalization scheme also allows for extracting higher excited state energies. Using the values of $c_3$ in Tab.~\ref{tab:Fit_c3} for various $\Lambda$s, we perform simulations on  
 $L^3 = 9^3, 10^3, \cdots, 14^3$ cubic lattice for the three-hadron system. The spectra of ground states and the first excited states are shown by 
  Fig.~\ref{fig:Efimov_state}. The ground states for the three $\Lambda$ values overlap with each other, as they are the fixed binding energy. 
  The standard angular momentum and parity projection technique~\cite{Lu:2014xfa} tells us that the quantum number of the ground state is $J^P=1^-$.
  The two energies above the ground state are the first excited state. The splitting of the first excited state is because there are two excitation modes for the three-body system. In the quark model language, the two modes are called $\rho$ and $\lambda$ excitation modes (for instance the discussion of Ref.~\cite{Zhong:2007gp}), with $\rho$ and $\lambda$ the two Jacobi coordinates. The detailed discussions can be found in the Appendix~\ref{app:C} and~\ref{app:D}. The splitting decreases with the increasing cubic volume $L^3$. From the figure, we can also see that, once the ground state is fixed, the first excited states for various cutoffs are close to each other. Their small differences are from the higher-order contribution, which is not implemented in our current framework. 
  This is due to the fact that our current frame work is renormalization group invariant~\cite{Lu:2023jyz} to leading order. 

\section{Summary}
\label{sec:Summary}

We first employ the novel lattice EFT method to multihadron systems. The most interesting system is the $DD^*K$ system with the existence of the experimental well-established states in the three two-body subsystems, i.e., the $T_{cc}^+$, $D^{\ast}_{s0}(2317)$, and $D_{s1}(2460)$. With the parameters of the two-body systems fixed by the binding energies of the two-body systems, we perform a calculation for the three-body system with a free parameter $c_3$ for the direct three-body interaction. When the three-body interaction is strongly attractive, the three-body system will collapse and is beyond the scope of our framework. When the three-body interaction is repulsive (even for the infinite repulsive interaction), the three-body energy is no larger than the $D_{s1}(2460)D$ threshold, making the three-body state existing unambiguously. For a $c_3\in (0,\infty)~\mathrm{MeV}^{-5}$, the three-body binding energy is located in the $(-84,-44)~\mathrm{MeV}$ region.    

To check the renormalization group invariance, we consider the binding energy with the parameter $c_3=0.1~\mathrm{MeV}^{-5}$ at $\Lambda=400~\mathrm{MeV}$ as an input to look at the binding energy of the first excited state. The results indicate that there is a splitting for the first excited states, which correspond to the $\rho$- and $\lambda$-type excitation. The splitting decreases with the increasing cubic size. In addition the standard angular momentum and parity projection technique is implemented for the quantum numbers of the ground and excited states. We find that both of them are $S$-wave states with quantum number $J^{P}=1^-$. Because the three-body state contains two charm quarks, it is easier to be detected in the Large Hadron Collider. Although this work is the first attempt to apply the novel lattice EFT to a multihadron system, it has exhibited the great power of lattice EFT to hadron physics, especially for many-body systems. 

\vspace{0.2cm}
{\bf \color{gray}Acknowledgements:}~~
We are grateful to Feng-Kun Guo, Kanchan P. Khemchandani, Peng-Yu Niu, Akaki Rusetsky, and Alberto Mart\'inez Torres for the helpful discussion. 
Q.W. would like to thank the long-term workshop on HHIQCD2024 at the Yukawa Institute for Theoretical Physics (YITP-T-24-02) for giving her a chance to deepen their ideas.
This work is partly supported by the National Natural Science Foundation of China with Grants No.~12375073, No.~12275259, and No.~12035007, NSAF No.U2330401, Guangdong Provincial funding with Grant No.~2019QN01X172, Guangdong Major Project of Basic and Applied Basic Research No.~2020B0301030008,
 the DFG (Project No. 196253076 - TRR 110), and the NSFC (Grant No. 11621131001) through the funds provided to the Sino-German CRC 110 “Symmetries and the Emergence of Structure in QCD”.

\nocite{*}
\bibliography{ref.bib}

\onecolumngrid
\newpage
\appendix

\clearpage
{\large \bf \section*{Supplemental Materials}}

\section{The details of the calculation on the lattice}

The particle masses used in the calculation (units in MeV) are
\begin{alignat*}{2}
     &M_{D^0} = 1864.84,\ \ \ \ &\ M_{D^+} = 1869.66,\ \ \ \ &\ M_D = 1867.25,\notag\\
     &M_{D^{\ast0}} = 2006.85,\ \ \ \ &\ M_{D^{\ast+}} = 2010.26,\ \ \ \ &\ M_{D^\ast}= 2008.56,\notag\\
     &M_{K^0} = 497.61,\ \ \ \ &\ M_{K^+} = 493.68,\ \ \ \ &\ M_{K}= 495.64,\notag\\
     &M_{\pi^0} = 134.98,\ \ \ \ &\ M_{\pi^\pm} = 139.57,\ \ \ \ &\ M_{\pi}= 138.04,\notag\\
     &M_{T_{cc}^+}= 3874.83,\ \ \ \ &\ M_{D^\ast_{s0}(2317)}= 2317.80,\ \ \ \ &\ M_{D_{s1}(2460)}= 2459.50,
\end{alignat*}
taken from Ref.~\cite{ParticleDataGroup:2008zun}.

For the two-body system, we perform simulations on $L^3=5^3,6^3,\cdots,19^3$ cubic lattices with a spatial lattice spacing $a=1/200\ \mathrm{MeV}^{-1}\approx0.99~\mathrm{fm}$. The cutoffs $\Lambda=400,~350,~300~\mathrm{MeV}$ are taken to check the renormalization group invariance. 
 For these calculations, we use the effective field theory Hamiltonian, i.e., Eq.~\eqref{eq:Hamiltonian_two_body}, and transform it from momentum space to coordinate space by a Fourier transform. To make a precision fit to the two-body energy, the extracted two-body parameters $v_0$, $h_3$, and $h_3^\ast$ from directly diagonalizing the lattice Hamiltonian using the Lanczos method are with little errors.
\begin{table}[htbp]
\renewcommand{\arraystretch}{1.2}
    \centering
    \caption{The parameter $v_0$ fitted to the $DD^\ast$ binding energy, on various $L^3$ cubic lattices with various cutoffs $\Lambda$.}
    \begin{tabular}{cp{2cm}<{\centering}p{3cm}<{\centering}p{3cm}<{\centering}p{3cm}<{\centering}}
    \hline
     Parameter (MeV$^{-2}$)&$L$&$\Lambda$=300(MeV)&$\Lambda$=350(MeV)&$\Lambda$=400(MeV) \\
    \hline
     \multirow{15}*{$v_0$}&5&-0.62&-0.58&-0.54\\
     & 6&-0.92&-0.83&-0.77\\
     & 7&-1.20&-1.07&-0.98\\
     & 8&-1.44&-1.28&-1.13\\
     & 9&-1.66&-1.43&-1.25\\
     & 10&-1.81&-1.54&-1.34\\
     & 11&-1.92&-1.62&-1.39\\
     & 12&-2.00&-1.67&-1.44\\
     & 13&-2.06&-1.71&-1.46\\
     & 14&-2.10&-1.74&-1.49\\
     & 15&-2.13&-1.76&-1.50\\
     & 16&-2.15&-1.78&-1.51\\
     & 17&-2.16&-1.79&-1.52\\
     & 18&-2.18&-1.79&-1.52\\
     & 19&-2.18&-1.80&-1.53\\
    \hline
    \end{tabular}  \label{III}
\end{table}
\clearpage

For the $D^{(*)}K$ interaction, there are three parameters, i.e. $h_1^{(*)}$, $h_3^{(*)}$, $h_5^{(*)}$. The redefined dimensionless low-energy constant $h_5^{\prime(*)}\equiv h_5^{(*)}M_D^2$ is set to $1$. The parameter $h_1^{(*)}=0.42$ is set to a constant.
\begin{table}[htbp]
\renewcommand{\arraystretch}{1.2}
    \centering
    \caption{The dimensionless parameter $h_3$ fitted to the $DK$ binding energy, on various $L^3$ cubic lattices with various cutoffs $\Lambda$.}
    \begin{tabular}{cp{2cm}<{\centering}p{3cm}<{\centering}p{3cm}<{\centering}p{3cm}<{\centering}}
    \hline
    Parameter&$L$&$\Lambda$=300(MeV)&$\Lambda$=350(MeV)&$\Lambda$=400(MeV) \\
    \hline
     \multirow{15}*{$h_3$}
     & 5 & -9.51 & -6.52 & -4.72 \\
     & 6 & -9.21 & -6.24 & -4.80 \\
     & 7 & -8.79 & -6.39 & -4.94 \\
     & 8 & -8.93 & -6.54 & -4.90 \\
     & 9 & -9.19 & -6.48 & -4.90 \\
     & 10& -9.15 & -6.46 & -4.91 \\
     & 11& -9.08 & -6.48 & -4.90 \\
     & 12& -9.09 & -6.48 & -4.90 \\
     & 13& -9.11 & -6.48 & -4.91 \\
     & 14& -9.11 & -6.48 & -4.91 \\
     & 15& -9.10 & -6.48 & -4.91 \\
     & 16& -9.10 & -6.48 & -4.91 \\
     & 17& -9.10 & -6.48 & -4.91 \\
     & 18& -9.10 & -6.48 & -4.91 \\
     & 19& -9.10 & -6.48 & -4.91 \\
    \hline
    \end{tabular}  \label{IV}
\end{table}

\begin{table}[htbp]
\renewcommand{\arraystretch}{1.2}
    \centering
    \caption{The dimensionless parameter $h_3^*$ fitted to the $D^*K$ binding energy, on various $L^3$ cubic lattices with various cutoffs $\Lambda$.}
    \begin{tabular}{cp{2cm}<{\centering}p{3cm}<{\centering}p{3cm}<{\centering}p{3cm}<{\centering}}
    \hline
     Parameter&$L$&$\Lambda$=300(MeV)&$\Lambda$=350(MeV)&$\Lambda$=400(MeV) \\
    \hline
     \multirow{15}*{$h^\ast_3$}
     & 5 & -9.97 & -6.78 & -4.86 \\
     & 6 & -9.65 & -6.48 & -4.94 \\
     & 7 & -9.20 & -6.64 & -5.10 \\
     & 8 & -9.35 & -6.80 & -5.05 \\
     & 9 & -9.63 & -6.73 & -5.05 \\
     & 10& -9.58 & -6.72 & -5.06 \\
     & 11& -9.51 & -6.74 & -5.06 \\
     & 12& -9.52 & -6.74 & -5.05 \\
     & 13& -9.54 & -6.73 & -5.06 \\
     & 14& -9.54 & -6.74 & -5.06 \\
     & 15& -9.53 & -6.74 & -5.06 \\
     & 16& -9.53 & -6.74 & -5.06 \\
     & 17& -9.53 & -6.74 & -5.06 \\
     & 18& -9.53 & -6.74 & -5.06 \\
     & 19& -9.53 & -6.74 & -5.06 \\
    \hline
    \end{tabular}  \label{tab:V}
\end{table}

With the values of the parameters $v_0$, $h_3$, and $h_3^\ast$ in the Tabs.~\ref{III}, \ref{IV}, and \ref{tab:V}, we perform simulations on $L^3=9^3,10^3,\cdots,14^3$ cubic lattice with a spatial lattice spacing $a=1/200\ \mathrm{MeV}^{-1}\approx0.99~\mathrm{MeV}$. The cutoff $\Lambda$ is also taken from 400 to 300 MeV with a constant interval of 50 MeV. After transforming the Hamiltonian of Eq.~\eqref{eq:Hamiltonian_three_body} from momentum space to coordinate space by a Fourier transform, one can extract the $DD^\ast K$ three-body binding energy from the matrix exact diagonalization scheme. Here, the dimensionless three-body interaction parameters $c_3=-1,~-0.1,~0,~0.1,~1,~10,~100~ \mathrm{MeV}^{-5}$ are used as an illustration.
\begin{table}[htbp]
\renewcommand{\arraystretch}{1.2}
    \centering
    \caption{The $DD^*K$ three-body binding energies with $c_3=-1,-0.1,0,0.1,1,10,100\ \mathrm{MeV}^{-5}$, cutoff $\Lambda=400,~350,~300~\mathrm{MeV}$ and $L^3=9^3, 10^3\cdots 14^3$ cubic lattice.}
    \begin{tabular}{ccp{2cm}<{\centering}p{2cm}<{\centering}p{2cm}<{\centering}p{2cm}<{\centering}p{2cm}<{\centering}p{2cm}<{\centering}c}
    \hline
     &$c_3$ (MeV$^{-5}$)&$L=9$&$L=10$&$L=11$&$L=12$&$L=13$&$L=14$&$\Lambda$ (MeV) \\
    \hline
     \multirow{7}*{$E_b$ (MeV)}
     & -1  & -113.81& -115.82& -115.56& -115.67& -116.14& -116.39&\multirow{7}*{300}\\
     & -0.1& -77.45 & -78.08 & -77.77 & -78.00 & -78.39 & -78.56 \\
     & 0   & -73.95 & -74.41 & -74.07 & -74.30 & -74.67 & -74.83 \\
     & 0.1 & -70.63 & -70.91 & -70.53 & -70.75 & -71.09 & -71.24 \\
     & 1   & -51.48 & -50.58 & -49.79 & -49.45 & -49.24 & -49.02 \\
     & 10  & -44.22 & -44.48 & -44.55 & -44.65 & -44.74 & -44.80 \\
     & 100 & -43.84 & -44.21 & -44.37 & -44.52 & -44.64 & -44.73 \\
    \hline
     \multirow{7}*{$E_b$ (MeV)}
     & -1  & -191.58& -191.08& -191.85& -192.49& -192.64& -192.84&\multirow{7}*{350}\\
     & -0.1& -86.79 & -87.00 & -87.70 & -88.11 & -88.28 & -88.47 \\
     & 0   & -77.36 & -77.54 & -78.18 & -78.54 & -78.70 & -78.88 \\
     & 0.1 & -68.93 & -69.01 & -69.56 & -69.85 & -69.98 & -70.14 \\
     & 1   & -46.47 & -46.09 & -45.86 & -45.68 & -45.54 & -45.45 \\
     & 10  & -44.67 & -44.72 & -44.79 & -44.84 & -44.87 & -44.90 \\
     & 100 & -44.56 & -44.64 & -44.74 & -44.80 & -44.84 & -44.88 \\
    \hline
     \multirow{7}*{$E_b$ (MeV)}
     & -1  & -357.63& -359.28& -359.96& -360.26& -360.66& -360.86&\multirow{7}*{400}\\
     & -0.1& -103.07& -104.17& -104.72& -105.10& -105.42& -105.62\\
     & 0   & -81.18 & -82.11 & -82.57 & -82.92 & -83.21 & -83.39 \\
     & 0.1 & -63.81 & -64.34 & -64.60 & -64.84 & -65.06 & -65.19 \\
     & 1   & -45.61 & -45.45 & -45.32 & -45.24 & -45.19 & -45.15 \\
     & 10  & -45.00 & -44.99 & -44.98 & -44.98 & -44.98 & -44.99 \\
     & 100 & -45.02 & -45.01 & -44.99 & -44.99 & -44.99 & -45.00 \\
    \hline
    \end{tabular}  \label{tab:three-body_binding_energy}
\end{table}
\clearpage
\section{The energy shift in the finite volume}\label{app:B}

The matrix exact diagonalization scheme uses the Implicitly Restarted Lanczos Method to find the eigenvalues and eigenvectors~\cite{lehoucq1998} by SciPy~\cite{Virtanen:2019joe} in Python. As this method becomes expensive for large volumes, we calculate the three-body binding energy $E_b$ in Tab.~\ref{tab:three-body_binding_energy} using small box sizes $L=10\to14$ and extrapolate the results to $L\to\infty$ by the energy shift function~\cite{Meissner:2014dea,Meng:2017jgx} in the finite volume,
\begin{align}
    \frac{\Delta E}{E_T}=&-(\kappa L)^{-3/2}\sum_{i=1}^{3}C_i\exp(-\mu_i\kappa L),
    \label{eq:volume_limit}
\end{align}
where $E_T$ and $\Delta E=E_T-E_L$ denote the binding energy in the infinite volume and the shift,
respectively. $\kappa=\sqrt{-2ME_T}$ is the bound-state momentum
with $M=(m_1+m_2+m_3)/6$ denoting a normalization mass, which inherits from Ref.~\cite{Nielsen:2001hbm} for the Jacobi coordinates of the three-body system. The value $M=(m_1+m_2+m_3)/6$ used in Ref.~\cite{Meng:2017jgx} and our work is for reproducing the result for the three identical particles in Ref.~\cite{Meissner:2014dea}. The factor \begin{align}
    \mu_{i}=&\sqrt{\frac{m_i(m_j+m_k)}{M(m_i+m_j+m_k)}}
\end{align}
and unknown coefficients $C_i$ depend on the masses in the system. 

Then, using non-linear least squares to fit the cost function $\chi^2=\sum_i (f(x_i)-y_i)^2$, where $f(x_i)$ is $E_L(L_i)$ in Eq.~\eqref{eq:volume_limit} and $y_i$ is the results in Tab.~\ref{tab:volume}.
\begin{table}[htbp]
\renewcommand{\arraystretch}{1.2}
    \centering
    \caption{The extracted three-body energy $E_b$ when $L\to\infty$ using Eq.~\eqref{eq:volume_limit}.}
    \begin{tabular}{p{2cm}<{\centering}p{2cm}<{\centering}p{5cm}<{\centering}}
    \hline
     $\Lambda$ (MeV)&$c_3$ (MeV$^{-5}$)&$E_T$ (MeV)\\
    \hline
     \multirow{7}*{300}
     & -1  & -116.40$\pm$ 0.05 \\
     & -0.1& -78.63 $\pm$ 0.006\\
     & 0   & -74.90 $\pm$ 0.001\\
     & 0.1 & -71.31 $\pm$ 0.005\\
     & 1   & -48.95 $\pm$ 0.07 \\
     & 10  & -44.83 $\pm$ 0.01 \\
     & 100 & -44.77 $\pm$ 0.01 \\
    \hline
     \multirow{7}*{350}
     & -1  & -192.77$\pm$0.10\\
     & -0.1& -88.46 $\pm$0.07\\
     & 0   & -78.88 $\pm$0.07\\
     & 0.1 & -70.15 $\pm$0.06\\
     & 1   & -45.39 $\pm$0.02\\
     & 10  & -44.92 $\pm$0.01\\
     & 100 & -44.90 $\pm$0.01\\
    \hline
     \multirow{7}*{400}
     & -1  & -360.46$\pm$0.47 \\
     & -0.1& -105.62$\pm$0.05 \\
     & 0   & -83.42 $\pm$0.04 \\
     & 0.1 & -65.24 $\pm$0.02 \\
     & 1   & -45.13 $\pm$0.01 \\
     & 10  & -44.99 $\pm$0.003\\
     & 100 & -45.00 $\pm$0.002\\
    \hline
    \end{tabular} \label{tab:volume}
\end{table}

\section{Angular momentum and parity projection}\label{app:C}

To probe the quantum numbers of the states of interest, 
the standard angular momentum and parity projection (AMP) technique is applied to the calculated eigenstate $|\Psi_0\rangle$. 
The essential tool in constructing the irreps is the angular momentum projection formula~\cite{Lu:2014xfa} is
\begin{align}
    |\Psi_A\rangle=&\frac{d_n}{24}\sum_{i=1}^{24}\chi_n(\Omega_i)R(\Omega_i)|\Psi_0\rangle,
\end{align}
where $|\Psi_0\rangle$ is a given eigenstate. Here $n=1,2,3,4,5$ corresponds to the five irreducible representations of the cubic rotational group, i.e. $A_1$, $A_2$, $E$,  $T_1$, and $T_2$. $d_n$ is the dimension of the corresponding irreducible representation. Their relation to the continuum SO(3) rotational group is presented in Tab.~\ref{tab:irreps}. $R(\Omega_i)$ is the $i$th rotational operator with $\Omega_i$ (with $i=1,2,\cdots 24$) the $i$th cubic rotational element, produced by the products of the $\pi/2$ rotation along the $x,y,z$ axes. $\chi_n$ is the character of the $n$th irreducible representation. We apply the AMP to all the ground and excited states in Fig.~\ref{fig:Efimov_state} and find that all of them are $S$-wave contributions, i.e. $A_1$. This demonstrates that both the ground and excited states in Fig.~\ref{fig:Efimov_state} have the quantum number $J^P=1^-$.  
\begin{table}[htbp]
\renewcommand{\arraystretch}{1.5}
    \centering
    \caption{Continuum total angular momenta decomposed into $irreps$(dim) of the $O$ group.}
    \begin{tabular}{p{2cm}<{\centering}p{5cm}<{\centering}}
    \hline
     SO(3)&$O$\\
    \hline
     J=0& $A_1(1)$\\
     J=1& $T_1(3)$\\
     J=2& $E(2)\oplus T_2(3)$\\
     J=3& $T_1(3)\oplus T_2(3)\oplus A_2(1)$\\
     J=4& $A_1(1)\oplus T_1(3)\oplus E(2)\oplus T_2(3)$\\
    \hline
    \end{tabular}  \label{tab:irreps}
\end{table}

\section{The splitting of the first excitation} \label{app:D}

\begin{figure}[htbp]
\centering
\subfigure[The ground state and excited states of the $DDD$ system with the two-body contact interaction strength $c_{DD}=-7~\mathrm{MeV}^{-2}$.]{\label{fig:test_1}
\includegraphics[width=0.43\textwidth]{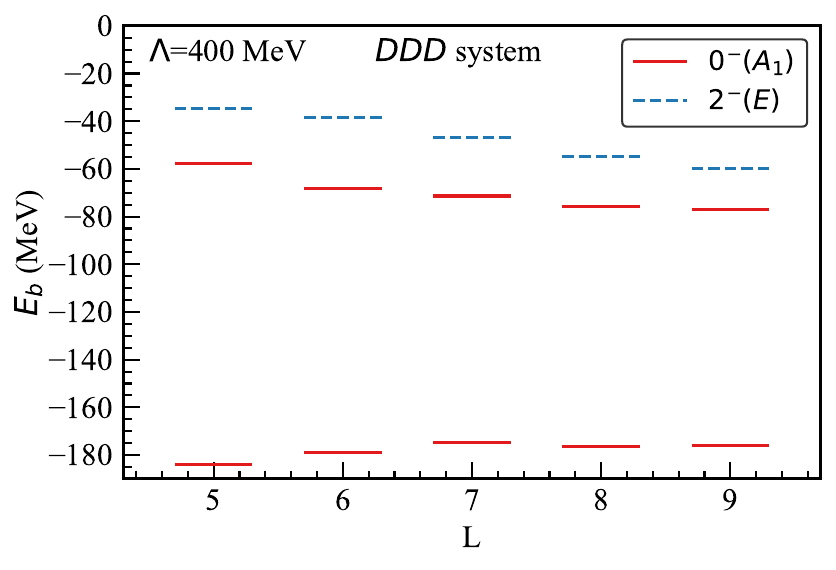}}
\subfigure[The ground state and the first excited states of the $DDK$ system with the two-body contact interaction strengths $c_{DD}=-1~\mathrm{MeV}^{-2}$ and $c_{DK}=-10~\mathrm{MeV}^{-2}$.]{\label{fig:test_2}
\includegraphics[width=0.43\textwidth]{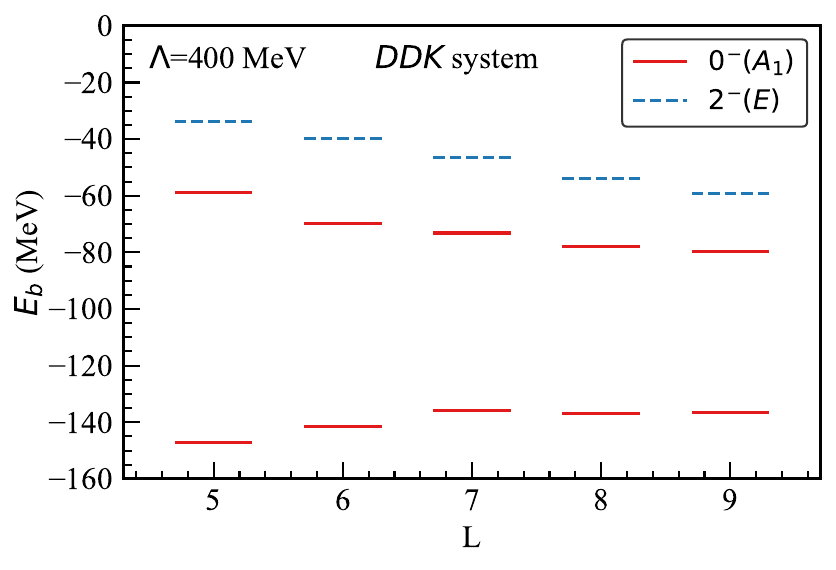}}
\vfill
\subfigure[The ground state and the first excited states of the $DDK$ system with the two-body contact interaction strengths $c_{DD}=-5~\mathrm{MeV}^{-2}$ and $c_{DK}=-8~\mathrm{MeV}^{-2}$.]{\label{fig:test_3}
\includegraphics[width=0.43\textwidth]{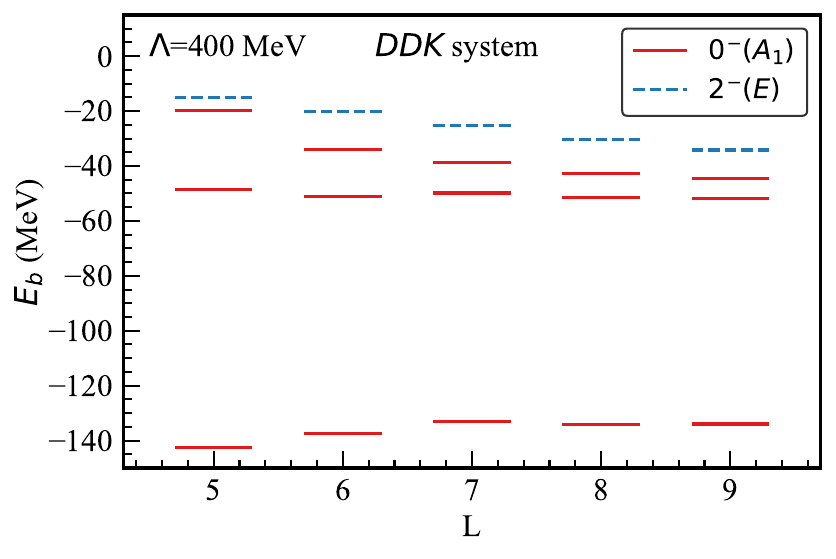}}
\subfigure[The ground state and the first excited states of the $DD^*K$ system with the two-body contact interaction strengths $c_{DD^*}=-1~\mathrm{MeV}^{-2}$, $c_{DK}=c_{D^*K}=-10~\mathrm{MeV}^{-2}$.]{\label{fig:test_4}
\includegraphics[width=0.43\textwidth]{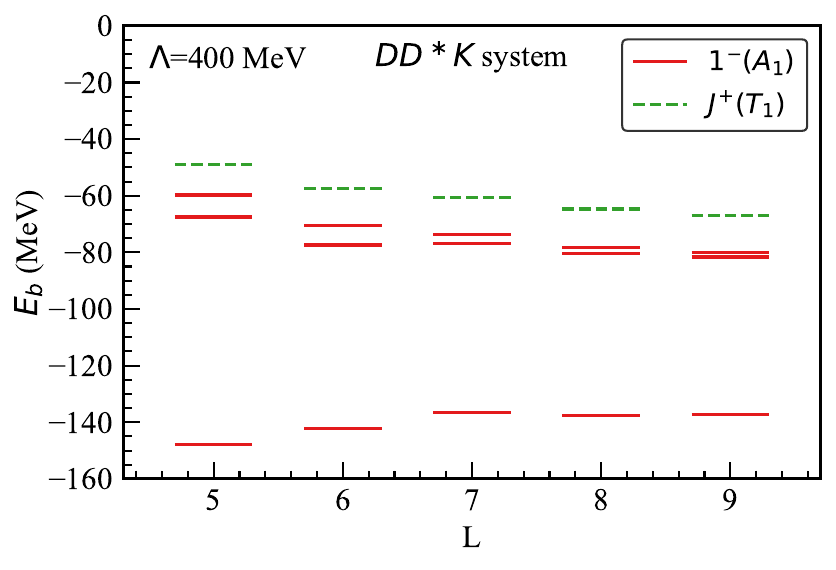}}
\caption{The $S$-wave ground (red solid curves) states, the first excited states (red solid curves), and the $D$-wave ground state (blue dashed curves) of the $DDD$ (a) and $DDK$ (b), (c) three-body systems. The red solid curves of the $DD^*K$ (d) three-body systems are for the $S$-wave ground and the first excited states. The green dashed curves in (d) are for the irreducible representation $T_1$ of the cubic rotational group with the angular momentum undetermined.}
\label{fig:}
\end{figure}

The splitting of the first excitation of the $DD^*K$ system is because there are two independent coordinates for the three-body system. More generally, 
in a three-body system,  $r_i$ is used to denote the coordinate for the $i$th particle in an arbitrary frame. One can transform them to the Jacobi coordinates,
\begin{align}
    \rho=&\frac{1}{\sqrt{2}}(r_1-r_2),\\
    \lambda=&\frac{1}{\sqrt{6}}(r_1+r_2-2r_3),\\
    R_{\mathrm{c.m.}}=&\frac{m_1r_1+m_2r_2+m_3r_3}{m_1+m_2+m_3}.
\end{align}
$R_\mathrm{c.m.}$ is the coordinate of the center-of-mass, which can be isolated out. The internal coordinates $\rho$ and $\lambda$ are independent coordinates and can be used in the spatial wave functions. Finally, a
 spatial wave function can be described as a product of the $\rho$- and $\lambda$-wave function. For instance, in Ref.~\cite{Zhong:2007gp}, the wave function of an oscillator is given by\begin{align}
    \psi_{l_\sigma m}^{n_\sigma}(\sigma)=&R_{n_\sigma l_\sigma}(\sigma)Y_{l_\sigma m}(\sigma),
\end{align}
where $\sigma=\rho,\lambda$. As the result, the first excited states have two excited modes, i.e. the $\rho$-type excitation and the $\lambda$-type excitation. Whether the two excited states degenerate or not depends on the masses and strengths of the three particles. In the following, we perform a zero-range (i.e. the $\delta$ potential) toy model to illustrate the above argument on $L^3=5^3,6^3,\cdots,9^3$ cubic lattice with the cutoff $\Lambda=400~\mathrm{MeV}$. The tested systems are set as follows:
\begin{itemize}
    \item[(i)] {For the identical particle system $DDD$, the two-body contact interaction strength is set as $c_{DD}=-7~\mathrm{MeV}^{-2}$. Because of their identity, the first excited states degenerate, as shown by Fig.~\ref{fig:test_1}.}
    \item[(ii)] {For the $DDK$ system with the two-body interaction strengths $c_{DD}=-1~\mathrm{MeV}^{-2}$ and $c_{DK}=-10~\mathrm{MeV}^{-2}$, because $c_{DD}\ll c_{DK}$, the two excited states still degenerate as shown in Fig.~\ref{fig:test_2},
although one particle mass becomes significantly smaller. }
    
    \item[(iii)] {In comparison with Fig.~\ref{fig:test_2}, for the $DDK$ system, when the two interaction strengths are comparable, for instance $c_{DD}=-5~\mathrm{MeV}^{-2}$ and $c_{DK}=-8~\mathrm{MeV}^{-2}$. The splitting of the two excited states emerge as shown by Fig.~\ref{fig:test_3}. }
     \item[(iv)] {In comparison with Fig.~\ref{fig:test_2}, when
     one $D$ meson is changed to one $D^*$ meson. The same two-body interaction strengths as those in Fig.~\ref{fig:test_2}, 
    i.e. $c_{DD^*}=-1~\mathrm{MeV}^{-2}$, $c_{DK}=c_{D^*K}=-10~\mathrm{MeV}^{-2}$, can also produce sizable splitting for the first excited states as shown in Fig.~\ref{fig:test_4}. }
\end{itemize}

\end{document}